%% file: main.tex
\documentclass[journal]{IEEEtran}
\usepackage{amsmath,amsfonts}
\usepackage{array}
\usepackage{textcomp}
\usepackage{stfloats}
\usepackage{url}
\usepackage{graphicx}
\usepackage{cite}
\usepackage{booktabs}
\usepackage{tabularx}
\newcommand{\ours}{\textsc{AgentWard}}

\begin{document}

\title{\ours{}: A Lifecycle Security Architecture for Autonomous AI Agents}

\author{Yixiang Zhang,
Xinhao Deng,
Jiaqing Wu,
Yue Xiao,
Ke Xu,
Qi Li
}


\maketitle

\begin{abstract}
\input{sections/0-abstract.tex}
\end{abstract}


\input{sections/1-introduction.tex}
\input{sections/2-background.tex}
\input{sections/3-architecture.tex}
\input{sections/4-evaluation.tex}
\input{sections/5-conclusion.tex}

\bibliographystyle{IEEEtran}
\bibliography{main}
\end{document}

%% file: sections/0-abstract.tex
Autonomous AI agents extend large language models into full runtime systems that load skills, ingest external content, maintain memory, plan multi-step actions, and invoke privileged tools. In such systems, security failures rarely remain confined to a single interface; instead, they can propagate across initialization, input processing, memory, decision-making, and execution, often becoming apparent only when harmful effects materialize in the environment.
This paper presents \ours{}, a lifecycle-oriented, defense-in-depth architecture that systematically organizes protection across these five stages. \ours{} integrates stage-specific, heterogeneous controls with cross-layer coordination, enabling threats to be intercepted along their propagation paths while safeguarding critical assets.
We detail the design rationale and architecture of five coordinated protection layers, and implement a plugin-native prototype on OpenClaw to demonstrate practical feasibility. This perspective provides a concrete blueprint for structuring runtime security controls, managing trust propagation, and enforcing execution containment in autonomous AI agents.

%% file: sections/1-introduction.tex
\section{Introduction}
\label{sec:introduction}

\IEEEPARstart{L}{arge} language model (LLM)-integrated systems have evolved from text-only question–answering chatbots to tool-augmented agents capable of short-term tasks, such as script generation and web search, and more recently to highly autonomous runtime systems that can delegate and execute end-to-end tasks, including project management and comprehensive research~\cite{yao2022react,  schick2023toolformer, openclaw2026}.
This progression is driven not only by advances in planning and tool-use capabilities of foundation models, but also by the evolution of agent architectures. Systems have evolved from simple tool-invocation pipelines to more sophisticated designs that support dynamic skill acquisition, long-term memory, and persistent workspaces, thereby enabling more complex and sustained agent behaviors~\cite{wu2023memgpt}.
OpenClaw~\cite{deng2026tamingopenclawsecurityanalysis,openclaw2026} exemplifies this trend. It supports community-contributed skills, maintains structured short- and long-term memory within a persistent workspace across sessions, and executes privileged tasks through controlled access to host tools under real runtime permissions.

This architectural and capability evolution broadens the security boundary from content safety to full runtime system security.  In traditional chatbot settings, security concerns primarily focus on unsafe inputs or outputs; with tool integration, risks can further arise from indirect prompt injection and tool-selection manipulation originating outside both the user and the model~\cite{greshake2023not, liu2023prompt, debenedetti2024agentdojo}.
Autonomous agents amplify these risks by introducing complex runtime modules, intermediate states, dynamic execution paths, and expanded privileges, which together enlarge the attack surface and increase the potential impact of failures. As a result, compromised skills, poisoned memory, or localized reasoning errors can escalate into file modification, service disruption, or data exfiltration~\cite{liu2026agent, dong2025minja, wei2025amemguard, andriushchenko2025agentharm}. These threats often propagate across the runtime pipeline: an attack may enter at one stage, persist through intermediate state, and manifest only later as harmful execution. Such patterns exceed the scope of pointwise defenses such as input filtering and call for end-to-end runtime security. For instance, a multi-step indirect prompt injection may evade input-level analysis, poison memory, and later trigger harmful actions, whereas memory integrity checks and behavior-level analysis can expose anomalous state and inconsistent execution patterns before the attack succeeds.

This article presents \ours{} as a lifecycle-oriented, defense-in-depth architecture for autonomous AI agents. We characterize the agent runtime lifecycle into five key stages: initialization, input, memory, decision, and execution. The framework organizes protection into five coordinated layers aligned with these stages, providing comprehensive coverage while disrupting attack propagation paths and safeguarding critical assets:
\begin{itemize}
    \item \textbf{Foundation Scan Layer}: Establishes a trusted baseline, focusing on the integrity of foundational system components.
    \item \textbf{Input Sanitization Layer}: Inspects runtime inputs, preventing the inbound entry and subsequent propagation of malicious data.
    \item \textbf{Cognition Protection Layer}: Safeguards the internal state, preventing the persistence and internal escalation of risks.
    \item \textbf{Decision Alignment Layer}: Constrains the inference process, limiting the propagation of risks from reasoning to action.
    \item \textbf{Execution Control Layer}: Governs environment-side effects, preventing the realization of harmful actions into external outcomes.
\end{itemize}
By spanning the full lifecycle with heterogeneous and complementary defense mechanisms, \ours{} mitigates complex, persistent, and evolving threats. Collectively, these layers isolate typical risk propagation paths and reinforce one another, thereby enhancing overall system robustness.


%% file: sections/2-background.tex
\section{A Lifecycle Security View of Threats}
\label{sec:threat-lifecycle}
Autonomous AI agents function as iterative systems designed for task completion. Triggered by an initial prompt, the agent leverages a large language model (LLM) to reason over external knowledge and operational guidance, maintains its internal state, and repeatedly invokes tools in a feedback-driven loop until the task is accomplished. This runtime process can be broken down into five distinct stages: initialization, input, memory, decision, and execution~\cite{deng2026tamingopenclawsecurityanalysis}.

At the initialization stage, the agent establishes its operating environment by loading necessary configurations, plugins, skills, and other capabilities. These components provide the external knowledge and tool interfaces required for performing complex tasks.
During the input stage, various inputs, such as prompts, documents, webpages, retrieval results, and tool responses, are incorporated into the current context.
In the memory stage, both short-term and long-term memory are managed. Short-term memory is maintained within the session and may be compressed when the dialogue exceeds the model's context window, while long-term memory is persistently stored and shared across sessions, retrieved as necessary. Additionally, user preferences and rules may be written into workspace files, such as \texttt{AGENTS.md} and \texttt{USER.md}.
The decision stage integrates available skills, callable tools, session history, memory, and the current user request into model inputs for reasoning, plan formulation, and tool selection.
Finally, during the execution stage, the agent translates its decisions into concrete actions, producing external side effects that complete the task.

\subsection{Threats to Agents Lifecycle}
These five lifecycle stages expose distinct security problems. At initialization, the key concern is whether the agent begins from a trustworthy baseline, since malicious plugins, poisoned skills, unsafe dependencies, or over-privileged configurations can compromise the capability set before any task begins~\cite{liu2026agent, ohm2020backstabber}. 
At input, the central problem is whether untrusted external content is admitted into the working context as passive data or misinterpreted as actionable control signals, enabling indirect prompt injection and related context-contamination attacks~\cite{greshake2023not, liu2023prompt, chen2025struq}. 
At memory, the risk shifts from transient influence to persistent corruption, because unsafe updates to session state or long-term memory can turn a local manipulation into durable behavioral bias~\cite{dong2025minja, srivastava2025memorygraft, wei2025amemguard}. 
At decision, the main problem is whether the agent's plan, tool choice, and parameters remain aligned with the authorized task~\cite{wallace2024instructionhierarchy, xie2025toolsafety}. 
At execution, those upstream deviations become concrete external effects, including unsafe command execution, unauthorized file or service modification, data exfiltration, and resource abuse~\cite{kuntz2025osharm, andriushchenko2025agentharm}.

\subsection{Cross-Stage Risk Propagation}
In autonomous agents, many real attacks do not remain local to the stage where they first appear, but instead evolve as cross-stage threat trajectories along the lifecycle. Common techniques such as prompt injection are therefore better understood as mechanisms for entering the runtime and altering intermediate state than as the final harm itself. Threats may originate from multiple points, including compromised initialization artifacts, untrusted external inputs, poisoned memory, or internally generated decision errors such as hallucination and objective drift. Once context, memory, or planning state is corrupted, these intermediate states become not only victims of attack but also relay points that propagate risk into later stages. For example, malicious content introduced at input may first poison memory, and the poisoned memory may later re-enter retrieval and context construction, thereby biasing subsequent decisions. As those decisions are translated into outputs and actions, the threat is further amplified into false user-facing responses, unauthorized plans, unsafe tool calls, or harmful execution outcomes. Autonomous-agent threats should therefore be understood as complex, persistent, and partially hidden paths of propagation across the lifecycle rather than as isolated attack instances tied to a single interface~\cite{greshake2023not, dong2025minja, srivastava2025memorygraft}.

%% file: sections/3-architecture.tex
\section{\ours{} Architecture}
\label{sec:architecture}

\begin{figure*}[]
\centering
\includegraphics[width=\textwidth]{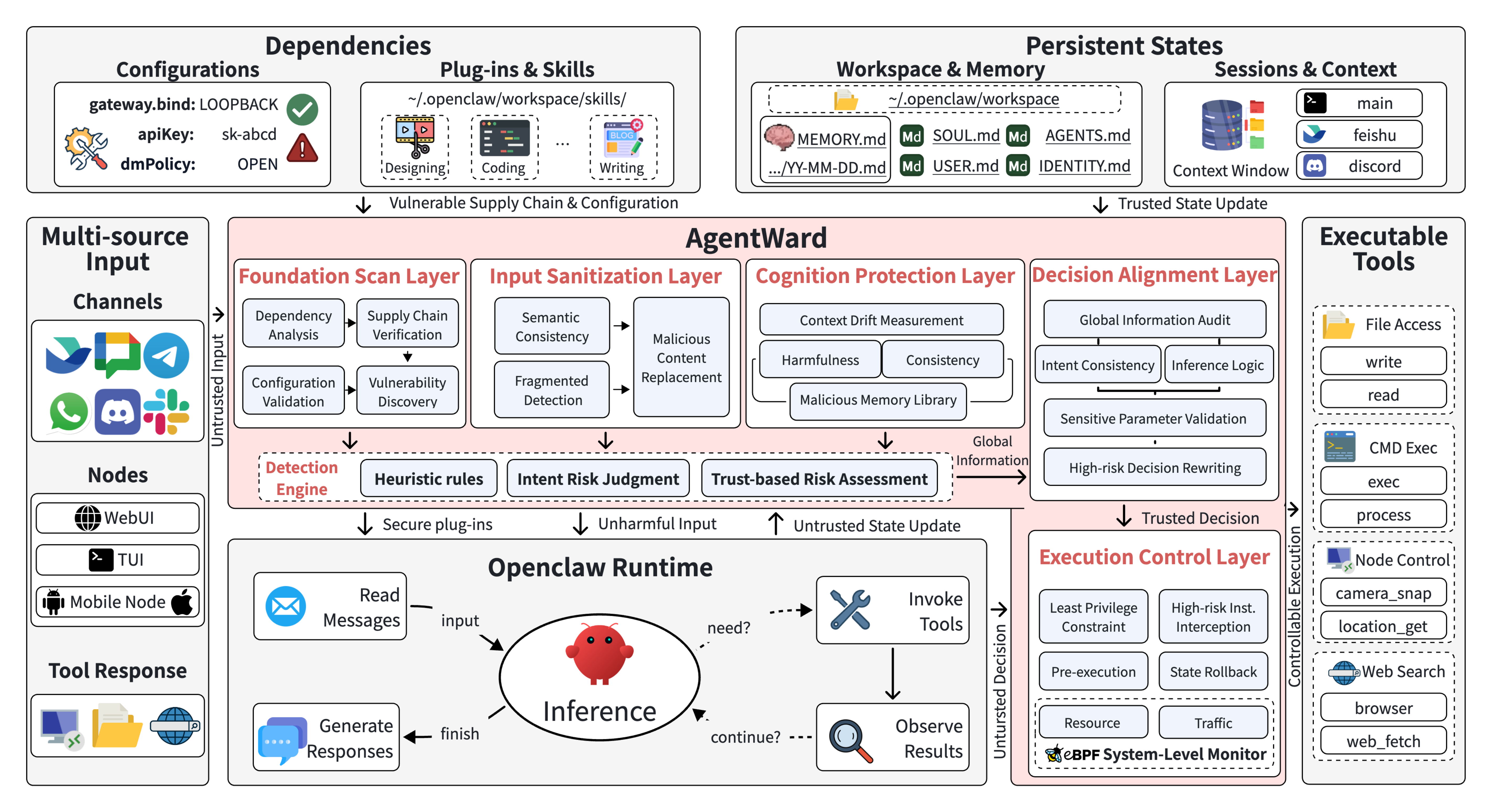}
\caption{Architectural overview of \ours{}. The framework attaches to lifecycle-relevant runtime events, organizes protection through five layers aligned with initialization, input, memory, decision, and execution, and carries security judgments forward through shared state and reusable analysis capabilities.}
\label{fig:overview}
\end{figure*}

In this section, we introduce \ours{}, a lifecycle-oriented runtime defense framework for autonomous AI agents. \ours{} adopts a layered defense-in-depth architecture that combines stage-specific heterogeneous controls with cross-layer coordination, enabling threats to be continuously detected, constrained, and interrupted as they propagate through the agent lifecycle. We first explain the rationale for full-lifecycle security and the resulting five-layer architecture, and then detail the role, boundary, inputs, outputs, and coordination of each layer. Figure~\ref{fig:overview} illustrates this organization in the OpenClaw runtime loop, where \ours{} is integrated into the lifecycle pipeline and provides security assurance through five coordinated layers: Foundation Scan, Input Sanitization, Cognition Protection, Decision Alignment, and Execution Control.

\subsection{Design Rationale}

The design of \ours{} is guided by four practical questions: 
What should the framework protect? 
What security posture should the framework adopt? 
How should the defenses be organized? 
These questions motivate four complementary design principles: full-lifecycle coverage, zero-trust enforcement, heterogeneous defense, and cross-layer coordination.

\textbf{Full-lifecycle Coverage}: 
\ours{} aims to protect the entire agent runtime system, including the agent, the assets it hosts, the environment in which it operates, and the users it serves. It also aims to interrupt plausible attack propagation paths before they result in external harm. To this end, \ours{} covers the full lifecycle of an autonomous agent, from initialization to execution. This design is necessary because threats against autonomous agents often cross stage boundaries rather than remaining confined to a single interface. Full-lifecycle coverage therefore reduces blind spots in the runtime pipeline and enables attacks to be intercepted at multiple points along their propagation paths.

\textbf{Zero-trust Enforcement}: 
\ours{} adopts a worst-case runtime posture: an attack may enter at any stage, contaminate any object, and propagate through any interface. Given the diversity and stealthiness of runtime attacks, each layer should assume that upstream components, retained state, and external signals may already be compromised or adversarially influenced. Accordingly, every layer evaluates local inputs, states, signals, and behaviors under a zero-trust posture, rather than relying solely on allow decisions made by preceding layers. As a result, an earlier allow decision does not automatically grant downstream passage, and later layers can still block, constrain, or escalate risky behavior under a fail-safe posture.

\textbf{Heterogeneous Defense}: 
Given the diversity of attack techniques and propagation paths, no single class of defense mechanism is sufficient for comprehensive protection. If all layers rely on similar detection logic or intervention strategies, an attacker may exploit a common bypass pattern that causes correlated failures across the framework. To improve robustness and reduce shared failure modes, \ours{} employs heterogeneous mechanisms across layers. These mechanisms are grounded in different defensive principles, protect different security objects, leverage different forms of evidence, and intervene in risky behavior through different means.

\textbf{Cross-layer Coordination}: 
Although each layer makes local security decisions, the layers should not operate in isolation. From a global perspective, an attack may propagate across multiple stages, leave weak signals in one stage, and only manifest harmful effects in another. A single-stage observation may appear suspicious but insufficient for a confident decision; by sharing context and accumulating risk evidence, \ours{} can connect dispersed signals into a more complete view of the attack, thereby improving detection accuracy and enabling more targeted responses. In addition, some layers may produce reusable analysis results, risk labels, or security states that can be inherited by subsequent layers. \ours{} therefore carries forward shared context, accumulated evidence, risk annotations, and reusable analysis outputs, allowing later stages to refine earlier findings, avoid redundant analysis, escalate responses progressively, and prevent conflicting interventions.

\subsection{Five-Layer Protection Design}
Following these principles, \ours{} organizes its defenses into five protection layers that collectively cover the full lifecycle of an autonomous agent, as shown in Table~\ref{tab:stage-interfaces}. Each layer protects a specific lifecycle stage, while shared security state and reusable analysis capabilities connect the layers into a cohesive defense system. The shared state carries risk evidence and annotations across stages, while reusable analysis capabilities provide common mechanisms for inspecting related objects, such as skills, inputs, memory, plans, and actions. This coordination allows evidence that is insufficient for one layer to be accumulated and reconsidered by later layers, and allows past detections to inform future assessments of similar attack surfaces. In this way, \ours{} forms a connected trust boundary that can detect, constrain, and interrupt threats as they propagate through the runtime lifecycle, rather than treating the lifecycle as a sequence of isolated checkpoints.
\input{tables/stage-interfaces-table.tex}

\subsubsection{Foundation Scan Layer}

The Foundation Scan Layer establishes the agent's trusted operating baseline at the beginning of the lifecycle. It operates at the runtime-supply chain boundary between the runtime and untrusted foundational components, including configurations, plugins, skills, dependencies, and permission declarations. This layer should inspect skills when they are introduced, upgraded, or modified, validate configurations to identify unsafe defaults or over-privileged settings, and examine plugins and dependencies for known vulnerabilities and integrity risks. It protects the runtime trust root and the approved capability baseline, because subsequent stages inherit the components, permissions, and default assumptions accepted during initialization.

The inputs to this layer include dependency metadata, supply-chain provenance, configuration files, plugin manifests, skill definitions, and permission declarations. Its outputs include initialization trust labels, warnings, policy constraints, and an approved capability baseline that later stages can use when forming local trust assumptions. As illustrated in Fig.~\ref{fig:overview}, a representative architecture first performs dependency analysis over both semantic descriptions and code-level dependencies to identify the components required by each skill at runtime. It then applies supply-chain verification to assess dependency integrity and trustworthiness, together with configuration validation to detect unsafe or vulnerable settings. The resulting evidence is aggregated for vulnerability discovery and baseline approval. These checks make initialization risks explicit and bounded before compromised components can influence context handling, memory updates, decision-making, or tool access~\cite{liu2026agent, ohm2020backstabber}.

\subsubsection{Input Sanitization Layer}

The Input Sanitization Layer governs the admission of information into the agent's active working context. It attaches to message ingress, retrieval results, document and webpage access, and tool-response ingestion, where external content from communication channels, knowledge sources, and environmental observations may begin to influence the model. This layer protects the instruction-data boundary and the active context consumed by subsequent memory and decision logic.

The inputs to this layer include multi-source external content, such as user-facing messages, retrieved passages, webpages, files, and tool outputs. Its outputs include provenance tags, quarantine decisions, restricted-use labels, sanitized or rewritten content, and input risk markers that can be reused by later stages. As illustrated in Fig.~\ref{fig:overview}, a representative design combines complementary checks, including semantic consistency analysis over input content, fragmented detection for split or distributed attacks, and multimodal consistency analysis for content spanning text, images, and other modalities. When unsafe segments are identified, these checks are followed by malicious-content replacement or rewriting. The role of this layer is to preserve useful information while ensuring that adversarial instructions are detected, downgraded, quarantined, or rewritten before they can contaminate downstream state, memory, or planning~\cite{greshake2023not, liu2023prompt, chen2025struq}.

\subsubsection{Cognition Protection Layer}

The Cognition Protection Layer governs the agent's context and persistent state. It monitors memory writes, session summaries, context merges, and updates to file-backed state. Its primary function is to ensure that changes to context or memory are valid, consistent, and free from contamination. This layer protects memory artifacts, workspace state, session context, and other durable state representations, because corruption in these objects can bias reasoning and behavior in subsequent tasks.

The inputs to this layer include state updates, memory payloads, session summaries, workspace writes, and state differences. Its outputs include trusted state updates, persistence approvals, memory read authorizations, rollback markers, downgrade-to-ephemeral decisions, and anomaly warnings. As shown in Fig.~\ref{fig:overview}, this layer checks for context drift, harmful memory content, and consistency violations. It may also use a library of malicious-memory patterns to detect known poisoning strategies. By controlling which state transitions become persistent and by governing memory reads, this layer prevents short-term contamination from evolving into long-term behavioral bias~\cite{dong2025minja, srivastava2025memorygraft, wei2025amemguard}.

\subsubsection{Decision Alignment Layer}

The Decision Alignment Layer reviews action formation before external side effects occur. It operates after the runtime has assembled a plan, selected tools, or generated execution-ready parameters, but before the proposed action is executed. At this stage, the action can still be revised, escalated for approval, or blocked. This layer protects intent-plan alignment and decision quality by checking whether the proposed action is justified by the user's authorized task, applicable security policies, and the accepted lifecycle context accumulated so far.

The inputs to this layer include user intent, accepted context, memory state, global information, and the proposed action. Its outputs include alignment scores, risk tiers, approval requirements, rewritten or blocked plans, and trusted-decision markers that can be inherited by the execution stage. As shown in Fig.~\ref{fig:overview}, a representative design consists of global information auditing, intent-consistency checking, inference-logic review, sensitive-parameter validation, and high-risk decision rewriting. This layer therefore serves as the primary semantic review point, where contaminated state, unsafe reasoning, or subtle goal drift can still be transformed into a controlled decision before any tool is invoked~\cite{xie2025toolsafety}.

\subsubsection{Execution Control Layer}

The Execution Control Layer governs the final agent-environment boundary, where an agent's internal decisions are translated into actions in the external environment. It attaches to tool calls, shell commands, file operations, API invocations, resource access, and network communication. This layer protects environment-facing assets, including files, services, devices, and network-accessible resources that may be affected by agent execution.

The inputs to this layer include concrete action requests, execution context, tool metadata, and upstream risk annotations produced by earlier layers. Its outputs include allow and deny decisions, least-privilege constraints, sandboxing policies, approval gates, rollback actions, kill signals, and controlled execution status. As shown in Fig.~\ref{fig:overview}, a representative design combines least-privilege enforcement, pre-execution checking, high-risk instruction interception, state rollback, resource monitoring, traffic monitoring, and an eBPF-based system monitor. By enforcing runtime constraints at the point where semantic risks can become concrete effects, this layer closes the lifecycle control loop and prevents harmful actions from being realized in the environment~\cite{owasp2025agentic, kuntz2025osharm, andriushchenko2025agentharm, xie2025toolsafety}.

\subsection{Cross-layer Coordination}

Cross-layer coordination is essential for defending autonomous agents against complex attack chains and stealthy multi-turn attacks. In \ours{}, each layer should not only make local security decisions, but also contribute its observations to a shared security context. This context records session-level risk evidence, trust annotations, and layer-specific judgments over key attack surfaces, including skills, memory, tools, tasks, and execution requests. \ours{} should also maintain an attack-path knowledge base that captures how risks enter the lifecycle, propagate across stages, and lead to harmful outcomes. This design allows the framework to reason about threats as evolving trajectories rather than isolated events.

This coordination serves two main purposes. First, it enables session-level risk accumulation. A single layer may observe weak or ambiguous evidence, such as suspicious input content, an uncertain memory update, or an unusual planning decision, but the evidence may be insufficient to justify immediate blocking. By carrying these signals forward, later layers can aggregate them with new observations and form a more reliable assessment. For example, if the Input Sanitization Layer marks a document as ambiguous but cannot confirm a prompt injection attempt, the Cognition Protection Layer and Decision Alignment Layer can reuse this marker when evaluating whether the content should be persisted, retrieved, or used to justify a high-risk action. The accumulated risk context can also be inherited by the Execution Control Layer, which may apply stricter execution policies, such as reduced tool permissions, additional approval gates, sandboxed execution, or fine-grained runtime monitoring. In this way, uncertain risks can be progressively escalated and translated into stronger downstream controls instead of being discarded at stage boundaries.

Second, cross-layer coordination enables history-aware defense adaptation. Since autonomous agents operate through iterative lifecycle loops across turns, tasks, and sessions, security judgments should not be treated as one-time decisions that disappear after a single stage or interaction. Instead, risk profiles and attack-path knowledge accumulated in previous lifecycle iterations can be reused to refine future assessments of similar behaviors. If a particular injection pattern, memory-poisoning strategy, or unsafe tool-use sequence has been observed before, the corresponding evidence should inform subsequent analysis and allow relevant layers to apply more conservative judgments. This adaptation can also be asymmetric across layers. An attack may bypass early semantic checks and memory validation, but later be intercepted by the Execution Control Layer when it attempts to trigger a high-risk operation. The resulting evidence should not remain local to execution; rather, it should be abstracted into reusable attack-path knowledge and propagated back to earlier layers, enabling earlier recognition of similar tactics in later lifecycle iterations.

By connecting layer-local judgments through shared security state and reusable attack knowledge, cross-layer coordination reduces the risk of isolated layer failures. It allows weak signals to accumulate within a session, enables prior experience to improve future defenses across lifecycle iterations, and allows heterogeneous layers to reinforce one another throughout the agent lifecycle.

%% file: tables/stage-interfaces-table.tex
\begin{table*}[t]
\caption{Lifecycle stages, protection layers, trust boundaries, and security objectives in \ours{}.}
\label{tab:stage-interfaces}
\centering
\small
\begin{tabular}{cccc}
\toprule
\textbf{Lifecycle Stage} & \textbf{Protection Layer} & \textbf{Trust Boundary} & \textbf{Security Objective} \\
\midrule
Initialization 
& Foundation Scan 
& Runtime--Supply Chain 
& Trusted operating baseline\\

Input 
& Input Sanitization 
& Instruction--Data 
& Active-context integrity \\

Memory 
& Cognition Protection 
& Transient--Persistent State 
& Internal-state integrity \\

Decision 
& Decision Alignment 
& Intent--Plan 
& Authorized plan formation \\

Execution 
& Execution Control 
& Agent--Environment
& Safe least-privilege execution \\
\bottomrule
\end{tabular}
\end{table*}

%% file: sections/4-evaluation.tex
\section{Implementation and Case Studies}
\label{sec:implementation-cases}
This section grounds \ours{} in a running prototype and then examines two attack chains that unfold across multiple lifecycle stages. The emphasis is on control placement, cross-layer coordination, and operational implication rather than on benchmark totals. Our goal is to show how the framework behaves when threats continue across initialization, context handling, memory mutation, planning, and execution.

\subsection{Prototype Implementation}

The current prototype implements \ours{} as a plugin-native extension to OpenClaw\footnote{Code is available at: \url{https://github.com/FIND-Lab/AgentWard}}~\cite{deng2026tamingopenclawsecurityanalysis}. Building on OpenClaw's plugin architecture, the prototype introduces a bidirectional plugin design. A unified adapter layer collects lifecycle-relevant runtime hooks and normalizes them into security events for the five protection layers, while layer-specific plugins consume these events and return structured outputs, including warning types, threat descriptions, judgment evidence, and block directives.
Concretely, Foundation Scan is attached to \texttt{before\_prompt\_build}. Input Sanitization and Decision Alignment are attached to \texttt{before\_message\_write}, and are applied to messages with the \texttt{tool} and \texttt{assistant} roles, respectively. Cognition Protection and Execution Control are attached to \texttt{before\_tool\_call}; the former is triggered only when memory files are about to be modified, whereas the latter monitors all tool calls. Per-session security state carries warnings and temporary blocks across turns, enabling later layers to inherit and act on earlier findings.
The current implementation covers malicious-skill and misconfiguration checks during initialization, tool-result contamination detection at input, memory-file modification anomaly detection at cognition, LLM-judge-based response review at decision, and pattern-based dangerous-command interception at execution. This scope is sufficient for the two case studies below, which aim to demonstrate layered interception and cross-stage coordination in a running agent system rather than to provide exhaustive benchmark coverage.

\subsection{Case Studies}

Rather than presenting one case per layer, we examine two lifecycle attack chains that cross multiple boundaries. This organization makes the framework's role clearer: \ours{} is most useful when an early finding constrains what later stages may store, plan, or execute, instead of remaining a local signal with no downstream effect.

\subsubsection{Case 1: Malicious Skill to Unauthorized Data Access}

The first case involves a user-installed skill whose description appears benign, while its files contain obfuscated instructions that steer the agent toward unauthorized credential access or data export~\cite{liu2026agent}. The attack first appears at the runtime-supply chain boundary, where the Foundation Scan Layer inspects the skill before the first reasoning turn. The layer detects a mismatch between the declared capability and the actual skill content, but the evidence is partially concealed and may be insufficient for immediate blocking. It therefore records the finding as a risk marker and propagates the warning to downstream layers.

Influenced by this skill, a benign programming task gradually shifts toward reading a sensitive file, packaging its contents, and preparing transmission to an external destination. At the decision stage, the Decision Alignment Layer compares the original user intent, the accepted context, the Foundation Scan warning, and the proposed action. This review identifies that the emerging plan exceeds the authorized task and conflicts with the user's intent. The accumulated risk is then inherited by the Execution Control Layer, which applies stricter tool permissions for the session. When the agent attempts to read the sensitive file or invoke a high-risk command, the action is blocked before it affects the environment.

This case illustrates that initialization risk cannot be handled as a one-time screening problem. A malicious skill enters as a capability artifact, but its harmful effect may only become visible after it influences planning or reaches execution. \ours{} preserves continuity across layers: an initialization warning lowers downstream trust, decision alignment exposes unauthorized plan drift, and execution control limits the remaining blast radius through final permission enforcement.

\subsubsection{Case 2: Indirect Prompt Injection to Persistent Backdoor and DoS}

The second case begins when a webpage or tool result contains an indirect prompt injection that appears sufficiently benign to enter the working context, especially under detection-only input filtering. This attack first crosses the instruction-data boundary. The Input Sanitization Layer examines returned content for prompt-template markers, jailbreak phrases, and dangerous instruction patterns. When suspicious content is identified, the layer can warn the agent, temporarily restrict tool-call permissions, or prevent contaminated tool results and derived responses from being persisted~\cite{wallace2024instructionhierarchy,chen2025struq}.
If the contaminated content still influences the session, the attack may escalate from transient context manipulation to persistent compromise. A typical next step is to write malicious instructions into durable state artifacts, such as \texttt{MEMORY.md}, \texttt{AGENTS.md}, or related workspace files. The Cognition Protection Layer inspects these state-changing operations at the tool boundary and blocks suspicious memory mutations, preventing the injected instructions from becoming trusted context in later sessions~\cite{dong2025minja, srivastava2025memorygraft, wei2025amemguard}. In this chain, the write-side intervention is critical because it prevents a future memory-loading step from reactivating the backdoor.

If this persistence path were left open, the poisoned memory could be retrieved in a later conversation and steer the agent toward harmful behavior, such as issuing commands that create a denial-of-service loop. \ours{} therefore constrains the attack at two temporal horizons. Cognition Protection blocks long-lived state corruption when persistence is attempted, while Execution Control blocks the final command pattern if a harmful action is still proposed at runtime. This case highlights the distinctive role of memory in lifecycle security: memory is both a target of poisoning and a relay point for future compromise. Protecting it therefore changes not only the current interaction, but also the attack surface of subsequent sessions.

The architectural lesson is that input filtering alone is insufficient when threats can persist across time. A lifecycle defense requires both a dedicated cognition boundary to prevent contaminated state from becoming durable, and a hard execution boundary to contain harmful reactivation if earlier defenses are bypassed.

%% file: sections/5-conclusion.tex
\section{Conclusion}
\label{sec:conclusion}
Autonomous agents should be secured as end-to-end systems with explicit runtime controls. A practical defense architecture starts from the observation that different lifecycle stages protect different assets and therefore require different control styles. Initialization and execution demand deterministic enforcement. Input, memory, and decision stages benefit from semantic reasoning, but only when the corresponding protection layers preserve their outputs as structured security state.

That is the central architectural lesson of \ours{}. Its value lies in combining a unified adapter layer for runtime events, five protection layers aligned with the lifecycle, reusable detection capabilities, and a small shared control vocabulary for warnings, policy flags, and intervention directives. If agent builders can map responsibilities to those layers, preserve provenance and trust labels across stages, restrict durable state transitions, validate plans before action, and enforce hard execution boundaries at the edge, they will be better positioned to deploy useful agents without assuming that any single model or guardrail will save them. The path to trustworthy autonomous agents is architectural clarity combined with coordinated lifecycle control.

%% file: main.bib
@misc{owasp2025agentic,
  author = {{OWASP Agentic Security Initiative}},
  title = {Agentic {AI} -- Threats and Mitigations},
  year = {2025},
  month = feb,
  url = {https://genai.owasp.org/resource/agentic-ai-threats-and-mitigations/},
  note = {Published Feb. 17, 2025. Accessed on Apr. 2, 2026}
}

@misc{liu2023prompt,
  author = {Liu, Yi and Deng, Gelei and Li, Yuekang and Wang, Kailong and Wang, Zihao and Wang, Xiaofeng and Zhang, Tianwei and Liu, Yepang and Wang, Haoyu and Zheng, Yan and Zhang, Leo Yu and Liu, Yang},
  title = {Prompt Injection Attack Against {LLM}-Integrated Applications},
  year = {2023},
  eprint = {2306.05499},
  archivePrefix = {arXiv},
  primaryClass = {cs.CR},
  doi = {10.48550/arXiv.2306.05499},
  url = {https://arxiv.org/abs/2306.05499},
  note = {Accessed on Apr. 2, 2026}
}

@misc{greshake2023not,
  author = {Greshake, Kai and Abdelnabi, Sahar and Mishra, Shailesh and Endres, Christoph and Holz, Thorsten and Fritz, Mario},
  title = {Not What You've Signed Up For: Compromising Real-World {LLM}-Integrated Applications with Indirect Prompt Injection},
  year = {2023},
  eprint = {2302.12173},
  archivePrefix = {arXiv},
  primaryClass = {cs.CR},
  url = {https://arxiv.org/abs/2302.12173},
  note = {Accessed on Apr. 16, 2026}
}

@misc{wallace2024instructionhierarchy,
  author = {Wallace, Eric and Xiao, Kai and Leike, Reimar and Weng, Lilian and Heidecke, Johannes and Beutel, Alex},
  title = {The Instruction Hierarchy: Training {LLM}s to Prioritize Privileged Instructions},
  year = {2024},
  eprint = {2404.13208},
  archivePrefix = {arXiv},
  primaryClass = {cs.CR},
  doi = {10.48550/arXiv.2404.13208},
  url = {https://arxiv.org/abs/2404.13208},
  note = {Accessed on Apr. 2, 2026}
}

@inproceedings{chen2025struq,
  author = {Chen, Sizhe and Piet, Julien and Sitawarin, Chawin and Wagner, David},
  title = {StruQ: Defending Against Prompt Injection with Structured Queries},
  booktitle = {34th {USENIX} Security Symposium ({USENIX Security} 25)},
  year = {2025},
  url = {https://www.usenix.org/conference/usenixsecurity25/presentation/chen-sizhe},
  note = {Accessed on Apr. 2, 2026}
}

@inproceedings{debenedetti2024agentdojo,
  author = {Debenedetti, Edoardo and Zhang, Jie and Balunovic, Mislav and Beurer-Kellner, Luca and Fischer, Marc and Tram{\`e}r, Florian},
  title = {AgentDojo: A Dynamic Environment to Evaluate Prompt Injection Attacks and Defenses for {LLM} Agents},
  booktitle = {Advances in Neural Information Processing Systems, Datasets and Benchmarks Track},
  year = {2024},
  url = {https://openreview.net/forum?id=m1YYAQjO3w},
  note = {NeurIPS 2024 Datasets and Benchmarks Track poster. Accessed on Apr. 2, 2026}
}

@inproceedings{andriushchenko2025agentharm,
  author = {Andriushchenko, Maksym and Souly, Alexandra and Dziemian, Mateusz and Duenas, Derek and Lin, Maxwell and Wang, Justin and Hendrycks, Dan and Zou, Andy and Kolter, J. Zico and Fredrikson, Matt and Gal, Yarin and Davies, Xander},
  title = {AgentHarm: A Benchmark for Measuring Harmfulness of {LLM} Agents},
  booktitle = {International Conference on Learning Representations},
  year = {2025},
  url = {https://openreview.net/forum?id=AC5n7xHuR1},
  note = {ICLR 2025 poster. Accessed on Apr. 2, 2026}
}

@inproceedings{kuntz2025osharm,
  author = {Kuntz, Thomas and Duzan, Agatha and Zhao, Hao and Croce, Francesco and Kolter, J. Zico and Flammarion, Nicolas and Andriushchenko, Maksym},
  title = {{OS}-{Harm}: A Benchmark for Measuring Safety of Computer Use Agents},
  booktitle = {Advances in Neural Information Processing Systems, Datasets and Benchmarks Track},
  year = {2025},
  url = {https://openreview.net/forum?id=Di30GwhQSX},
  note = {NeurIPS 2025 Datasets and Benchmarks Track spotlight. Accessed on Apr. 9, 2026}
}

@misc{openclaw2026,
  author = {Peter Steinberger and the OpenClaw contributors},
  title = {{OpenClaw}: Personal {AI} Assistant},
  year = {2026},
  url = {https://github.com/openclaw/openclaw},
  note = {GitHub repository. Accessed on Apr. 16, 2026}
}

@misc{liu2026agent,
  author = {Liu, Yi and Wang, Weizhe and Feng, Ruitao and Zhang, Yao and Xu, Guangquan and Deng, Gelei and Li, Yuekang and Zhang, Leo},
  title = {Agent Skills in the Wild: An Empirical Study of Security Vulnerabilities at Scale},
  year = {2026},
  eprint = {2601.10338},
  archivePrefix = {arXiv},
  primaryClass = {cs.CR},
  url = {https://arxiv.org/abs/2601.10338},
  note = {Accessed on Apr. 16, 2026}
}

@misc{wei2025amemguard,
  author = {Wei, Qianshan and Yang, Tengchao and Wang, Yaochen and Li, Xinfeng and Li, Lijun and Yin, Zhenfei and Zhan, Yi and Holz, Thorsten and Lin, Zhiqiang and Wang, XiaoFeng},
  title = {A-MemGuard: A Proactive Defense Framework for {LLM}-Based Agent Memory},
  year = {2025},
  eprint = {2510.02373},
  archivePrefix = {arXiv},
  primaryClass = {cs.CR},
  doi = {10.48550/arXiv.2510.02373},
  url = {https://arxiv.org/abs/2510.02373},
  note = {Accessed on Apr. 27, 2026}
}

@inproceedings{xie2025toolsafety,
  author = {Xie, Yuejin and Yuan, Youliang and Wang, Wenxuan and Mo, Fan and Guo, Jianmin and He, Pinjia},
  title = {ToolSafety: A Comprehensive Dataset for Enhancing Safety in {LLM}-Based Agent Tool Invocations},
  booktitle = {Proceedings of the 2025 Conference on Empirical Methods in Natural Language Processing},
  year = {2025},
  doi = {10.18653/v1/2025.emnlp-main.714},
  url = {https://aclanthology.org/2025.emnlp-main.714/},
  note = {Accessed on Apr. 12, 2026}
}

@misc{deng2026tamingopenclawsecurityanalysis,
      title={Taming OpenClaw: Security Analysis and Mitigation of Autonomous LLM Agent Threats}, 
      author={Xinhao Deng and Yixiang Zhang and Jiaqing Wu and Jiaqi Bai and Sibo Yi and Zhuoheng Zou and Yue Xiao and Rennai Qiu and Jianan Ma and Jialuo Chen and Xiaohu Du and Xiaofang Yang and Shiwen Cui and Changhua Meng and Weiqiang Wang and Jiaxing Song and Ke Xu and Qi Li},
      year={2026},
      eprint={2603.11619},
      archivePrefix={arXiv},
      primaryClass={cs.CR},
      url={https://arxiv.org/abs/2603.11619}, 
}

@misc{ohm2020backstabber,
  author = {Ohm, Marc and Plate, Henrik and Sykosch, Arnold and Meier, Michael},
  title = {Backstabber's Knife Collection: A Review of Open Source Software Supply Chain Attacks},
  year = {2020},
  eprint = {2005.09535},
  archivePrefix = {arXiv},
  primaryClass = {cs.CR},
  doi = {10.48550/arXiv.2005.09535},
  url = {https://arxiv.org/abs/2005.09535},
  note = {Accessed on Apr. 27, 2026}
}

@misc{yao2022react,
  author = {Yao, Shunyu and Zhao, Jeffrey and Yu, Dian and Du, Nan and Shafran, Izhak and Narasimhan, Karthik and Cao, Yuan},
  title = {ReAct: Synergizing Reasoning and Acting in Language Models},
  year = {2022},
  eprint = {2210.03629},
  archivePrefix = {arXiv},
  primaryClass = {cs.CL},
  doi = {10.48550/arXiv.2210.03629},
  url = {https://arxiv.org/abs/2210.03629},
  note = {Accessed on Apr. 27, 2026}
}

@misc{schick2023toolformer,
  author = {Schick, Timo and Dwivedi-Yu, Jane and Dess{\`\i}, Roberto and Raileanu, Roberta and Lomeli, Maria and Zettlemoyer, Luke and Cancedda, Nicola and Scialom, Thomas},
  title = {Toolformer: Language Models Can Teach Themselves to Use Tools},
  year = {2023},
  eprint = {2302.04761},
  archivePrefix = {arXiv},
  primaryClass = {cs.CL},
  doi = {10.48550/arXiv.2302.04761},
  url = {https://arxiv.org/abs/2302.04761},
  note = {Accessed on Apr. 27, 2026}
}

@misc{wu2023memgpt,
  author = {Packer, Charles and Wooders, Sarah and Lin, Kevin and Fang, Vivian and Patil, Shishir G. and Stoica, Ion and Gonzalez, Joseph E.},
  title = {{MemGPT}: Towards {LLM}s as Operating Systems},
  year = {2023},
  eprint = {2310.08560},
  archivePrefix = {arXiv},
  primaryClass = {cs.AI},
  doi = {10.48550/arXiv.2310.08560},
  url = {https://arxiv.org/abs/2310.08560},
  note = {Accessed on Apr. 27, 2026}
}

@misc{dong2025minja,
  author = {Dong, Shen and Xu, Shaochen and He, Pengfei and Li, Yige and Tang, Jiliang and Liu, Tianming and Liu, Hui and Xiang, Zhen},
  title = {Memory Injection Attacks on {LLM} Agents via Query-Only Interaction},
  year = {2025},
  eprint = {2503.03704},
  archivePrefix = {arXiv},
  primaryClass = {cs.LG},
  doi = {10.48550/arXiv.2503.03704},
  url = {https://arxiv.org/abs/2503.03704},
  note = {Accessed on Apr. 27, 2026}
}

@misc{srivastava2025memorygraft,
  author = {Srivastava, Saksham Sahai and He, Haoyu},
  title = {MemoryGraft: Persistent Compromise of {LLM} Agents via Poisoned Experience Retrieval},
  year = {2025},
  eprint = {2512.16962},
  archivePrefix = {arXiv},
  primaryClass = {cs.CR},
  doi = {10.48550/arXiv.2512.16962},
  url = {https://arxiv.org/abs/2512.16962},
  note = {Accessed on Apr. 27, 2026}
}
